\documentclass[journal]{IEEEtran}

\usepackage{cite}
\usepackage{graphicx}
\usepackage{amsmath}
\usepackage{amsfonts}

\usepackage{cite}
\usepackage{graphicx}
\graphicspath{{Figures/}}

\usepackage{amsmath}
\usepackage{algorithmic}
\usepackage{array}
\usepackage[caption=false,font=footnotesize]{subfig}
\usepackage{url}
\usepackage{orcidlink}
\usepackage{booktabs}
\usepackage{multirow}
\usepackage{xcolor}
\usepackage{makecell}

\begin{document}

\title{\title{RFM-Editing 2: Text-Guided Audio Editing with Rectified Flow Matching and Coarse-to-Fine Diffusion Transformers}}

\author{
Liting~Gao\,\orcidlink{0009-0004-6659-382X},
Yonggang~Zhu\,\orcidlink{0009-0003-5933-8049},
Yaru~Chen\,\orcidlink{0009-0001-6648-0399},
Dongyu~Wang\,\orcidlink{0009-0008-4288-2714},
Shubin~Zhang\,\orcidlink{0000-0002-7985-4898},~\IEEEmembership{Member,~IEEE},\\
Zhenbo~Li\,\orcidlink{0000-0003-2914-1192},
Jean-Yves~Guillemaut\,\orcidlink{0000-0001-8223-5505},
and Wenwu~Wang*\,\orcidlink{0000-0002-8393-5703},~\IEEEmembership{Fellow,~IEEE}
\thanks{Liting Gao, Yaru Chen, Dongyu Wang, Jean-Yves Guillemaut, and Wenwu Wang are with the Centre for Vision, Speech and Signal Processing (CVSSP), University of Surrey, Guildford GU2 7XH, U.K. E-mail: \{l.gao, yaru.chen, dongyu.wang, j.guillemaut, w.wang\}@surrey.ac.uk.}%
\thanks{Yonggang Zhu is with the School of Artificial Intelligence, Beijing University of Posts and Telecommunications, Beijing, China. E-mail: zhuyonggang@bupt.edu.cn.}%
\thanks{Shubin Zhang is with the Fisheries College, Ocean University of China, Qingdao, China. E-mail: zhangshubin@ouc.edu.cn.}%
\thanks{Zhenbo Li is with the College of Information and Electrical Engineering, China Agricultural University, Beijing, China. E-mail: lizb@cau.edu.cn.}%
\thanks{*Corresponding author.}%
}

%\markboth{Journal of \LaTeX\ Class Files,~Vol.~14, No.~8, April~2026}%

\maketitle

\begin{abstract}
Audio editing aims to modify specific content in an existing audio clip according to a text instruction or description while preserving the remaining acoustic content. Despite the remarkable progress of diffusion models, existing training-based editing methods mainly rely on the local inductive biases and cross-attention interaction in convolutional U-Net backbones, which often hinder long-range semantic alignment and precise understanding and localization of instructions.
In contrast, diffusion transformers provide stronger global modeling and multimodal fusion, but existing editing architectures usually adopt a simple stack of diffusion transformer blocks. Applying joint attention over concatenated audio and text tokens in all blocks results in quadratic complexity with respect to token length.
To balance editing performance and efficiency, we propose a novel instruction-guided audio editing framework based on rectified flow matching (RFM), named RFM-Editing 2, built on a hybrid two-stage diffusion transformer. The proposed model performs joint attention over audio and text tokens to establish coarse semantic alignment at the low-resolution stage, then switches to alternating joint-attention and cross-attention blocks to refine editing details at the high-resolution stage. This coarse-to-fine strategy enables efficient and accurate instruction-guided audio editing. Experiments show that the proposed framework achieves notable performance gains on challenging editing tasks involving overlapping audio events and complex instructions, while substantially improving editing efficiency.
\end{abstract}


\IEEEpeerreviewmaketitle

\section{Introduction}
\IEEEPARstart{G}{enerative} diffusion models have led to remarkable progress in text-to-audio (TTA) synthesis, with representative approaches including denoising diffusion probabilistic model (DDPM)-based methods \cite{ho2020denoising}, such as AudioLDM \cite{liu2023audioldm, liu2024audioldm}, Make-An-Audio \cite{huang2023make, huang2023make2}, Tango \cite{ghosal2023text, majumder2024tango}, and flow-based methods \cite{lipman2022flow, liu2022flow}, such as TangoFlux \cite{hung2024tangoflux}. By leveraging large-scale audio-text pretraining and powerful generative backbones, these modern TTA systems can synthesize high-fidelity, semantically aligned, and diverse audio directly from natural language. This progress in generative modeling and TTA has also stimulated growing interest in controllable audio editing in recent years.
% enabled a paradigm shift from text-to-audio generation from scratch to controllable audio editing in recent years.

Text-guided audio editing \cite{wang2023audit, manor2024zero} aims to modify existing audio according to text instructions or descriptions while faithfully preserving non-target content. In real-world scenarios, users often seek to modify specific sound events within an audio clip, making precise localization and efficient editing particularly important. As a result, audio editing demands not only generative capability, but also fine-grained control and content preservation in continuous temporal signals where multiple sound events may overlap. Text-guided audio editing enables intuitive and flexible audio manipulation without manual waveform editing or expert post-processing, supporting applications such as sound design, post-production, and personalized audio creation.

Existing text-guided audio editing works can be broadly categorized into training-free \cite{jia2025audioeditor, xu2024prompt, manor2024zero} and training-based methods \cite{wang2023audit, paissan2024audio, gao2026rfm, tao2025mmedit, zhu2026audio}. Training-free methods often rely on costly inversion or optimization at inference time and require detailed captions rather than concise editing instructions. Training-based methods enable explicit instruction following through supervision, yet they often struggle in complex acoustic scenarios with multiple overlapping sound events. A key reason is that convolutional U-Net editors \cite{wang2023audit, gao2026rfm} mainly rely on text--audio cross-attention interaction and strong local inductive biases, which can limit long-range semantic alignment as well as precise instruction understanding and localization.

Although transformer-based diffusion methods provide stronger global modeling and multimodal fusion, existing models \cite{tao2025mmedit, zhu2026audio} typically stack multimodal diffusion transformer (MMDiT) \cite{esser2024scaling} and diffusion transformer (DiT) blocks uniformly under the Flux-style \cite{flux2024, labs2025flux1kontextflowmatching} architecture. We argue that coarse semantic alignment and fine acoustic refinement require different multimodal interaction mechanisms. Moreover, applying joint attention over fully concatenated audio and text tokens in all blocks introduces quadratic complexity with respect to token length, making it difficult to balance editing accuracy and efficiency.

To address these limitations, we propose a new framework, named RFM-Editing 2, for text-guided audio editing based on continuous-time rectified flow matching (RFM) and a two-stage hybrid diffusion transformer. Global conditioning information modulates the normalized audio tokens through adaptive layer normalization (AdaLN), while token-level audio--text interactions are modeled through attention mechanisms. Our model adopts a coarse-to-fine design, where joint attention is performed over audio and text features at the low-resolution stage to establish coarse semantic alignment, while local editing details are refined at the high-resolution stage by switching to the alternation between joint-attention and cross-attention blocks. In this way, the proposed framework better balances global semantic understanding, local editing precision, and inference efficiency.

This work extends our preliminary work, RFM-Editing~\cite{gao2026rfm}. While RFM-Editing first formulated instruction-guided audio editing under a U-Net-based RFM framework, this work further improves editing precision and inference efficiency through the proposed two-stage hybrid diffusion transformer architecture. The coarse-to-fine framework uses dual-stream joint-attention MMDiT blocks for global audio--text fusion at the low-resolution stage, followed by alternating DiT and MMDiT blocks for high-resolution acoustic refinement. We further introduce global AdaLN-Zero modulation by jointly conditioning on the timestep, instruction, and original audio. In addition, we expand the training data by constructing audio editing datasets from AudioCaps, AudioSet, and AudioSetCaps. Experiments provide a more complete evaluation with task-wise comparative results against representative baselines, together with systematic ablation studies.

%The coarse-to-fine framework uses dual-stream joint-attention MMDiT blocks for global audio--text fusion at the low-resolution stage, followed by alternating DiT and MMDiT blocks for high-resolution acoustic refinement. We further introduce global AdaLN-Zero modulation by jointly conditioning on the timestep, instruction, and original audio. In addition, we expand the training data by constructing audio editing datasets from AudioCaps, AudioSet, and AudioSetCaps.

Our contributions can be summarized as follows:
\begin{itemize}
    \item We propose a novel framework for text-guided audio editing based on continuous-time rectified flow matching in the latent space and a two-stage hybrid diffusion transformer, which improves editing performance while substantially reducing inference time.

    \item We design a hierarchical architecture in which the Dual-Stream Joint-Attention MMDiT (DSJA-MMDiT) blocks are used at the low-resolution stage for coarse audio--text semantic fusion, while the DSJA-MMDiT and the AdaLN-Zero Cross-Attention DiT (AZCA-DiT) blocks are alternately stacked at the high-resolution stage for fine-grained audio refinement and local acoustic detail recovery. This coarse-to-fine design helps balance cross-modal interaction and inference efficiency.

    \item We fuse global text and original-audio features with timestep embeddings for AdaLN-Zero modulation, while using the token-level text and original-audio features for fine-grained conditioning, thereby improving editing controllability and the preservation of non-edited content.
\end{itemize}

The remainder of this paper is organized as follows. Section II reviews related work on traditional, training-free, and training-based audio editing methods. Section III presents the proposed RFM-Editing 2 framework. Section IV describes the experimental setup and discusses the results. Finally, Section V concludes the paper.

\section{Related Work}
\label{sec:related}

This section reviews traditional signal-processing-based audio editing methods that predate modern generative models. We then focus on modern diffusion- and flow-based audio editing methods, which can be broadly categorized into training-based and training-free paradigms. This work primarily focuses on the training-based paradigm.

\subsection{Traditional Audio Editing Methods}
Traditional audio editing mainly relied on signal processing with time-frequency representations such as short-time Fourier transform (STFT)~\cite{allen2005unified}, followed by spectral modification and waveform reconstruction. Classical techniques include phase vocoder methods~\cite{dolson1986phase, laroche1999improved} for time-scale and pitch manipulation and Griffin--Lim reconstruction~\cite{griffin1984signal} for waveform recovery from modified magnitude spectrograms. Beyond spectral editing, some methods formulated audio restoration and inpainting as inverse problems with handcrafted priors, such as sparse reconstruction in transform domains~\cite{adler2011audio}. Spectral modeling synthesis~\cite{serra1990spectral} enabled interpretable manipulation through signal decomposition and parametric modeling. The speech-specific waveform editing method PSOLA~\cite{moulines1990pitch} enabled prosody modification by manipulating pitch-synchronous segments.
Although these methods are mathematically interpretable and computationally efficient, they mainly operate at the signal level and depend on handcrafted representations or restrictive audio structure assumptions. Thus, they are less suitable for complex acoustic scenes and high-level semantic-driven editing, motivating the shift toward data-driven generative audio editing models.

\subsection{Modern Audio Editing Methods}
\subsubsection{\textbf{Diffusion and Flow Matching Models}}
Diffusion models~\cite{ho2020denoising, song2021denoising} and flow matching models~\cite{lipman2022flow, liu2022flow} have demonstrated strong performance in modern audio editing~\cite{wang2023audit, manor2024zero, jia2025audioeditor, gao2026rfm} by modeling complex data distributions through iterative denoising or continuous-time transport.

DDPMs \cite{ho2020denoising} define a Markovian forward diffusion process $q(x_t \mid x_{t-1})=\mathcal{N}(x_t; \sqrt{1-\beta_t} x_{t-1}, \beta_t I)$, which gradually corrupts the clean data sample $x_0$ into Gaussian noise. The forward noising process can be reparameterized as
\begin{equation}
x_t
=
\sqrt{\bar{\alpha}_t}\,x_0
+
\sqrt{1-\bar{\alpha}_t}\,\boldsymbol{\epsilon},
\qquad
\boldsymbol{\epsilon} \sim \mathcal{N}(\mathbf{0}, \mathbf{I}),
\end{equation}
where $\beta_t$ is the noise variance schedule at timestep $t$, and $\bar{\alpha}_t=\prod_{i=1}^t(1-\beta_i)$ is the cumulative signal retention coefficient. A denoising network $\boldsymbol{\epsilon}_\theta(x_t, t)$, typically implemented with U-Net \cite{ronneberger2015u} or, more recently, transformer-based DiT \cite{peebles2023scalable}, is trained to predict the added noise using a mean-squared error (MSE) objective. Generation is performed by gradually reversing the diffusion process from noise to data. 

% DDPMs have been successfully extended to audio generation \cite{kong2020diffwave, yang2023diffsound}.
% To improve scalability for high-dimensional data, latent diffusion models (LDMs) \cite{rombach2022high} perform diffusion in a compact latent space learned by a pre-trained autoencoder \cite{kingma2013auto}. Combined with the classifier-free guidance (CFG) \cite{ho2022classifier}, they form the basis of modern TTA systems \cite{liu2023audioldm, liu2024audioldm, huang2023make, huang2023make2, ghosal2023text}. 
Beyond discrete-time formulations, diffusion models can be reformulated as continuous-time stochastic differential equations (SDEs) \cite{song2021scorebased}, which further implies the existence of a deterministic ordinary differential equation (ODE) known as the Probability Flow ODE \cite{song2021scorebased}. This perspective provides a theoretical foundation for deterministic trajectory-based sampling in DDIM \cite{song2021denoising} and stable inversion \cite{mokady2023null, huberman2024edit} used in audio editing \cite{manor2024zero, jia2025audioeditor}. 
% It also connects to consistency models~\cite{song2023consistency}, which enable efficient few-step generation.

Flow Matching (FM) \cite{lipman2022flow} learns a time-dependent vector field that transports samples along a prescribed probability path in continuous time, instead of predicting noise.
Given samples ${x}_t$ with target velocity ${v}_t$, FM minimizes
\begin{equation}
\mathcal{L}_{\text{FM}}
=
\mathbb{E}_{{x}_t,\,t}
\bigl[
\lVert
{v}_\theta({x}_t,t)-{v}_t
\rVert_2^2
\bigr].
\end{equation}
This objective avoids explicit likelihood computation and discrete noise schedules. Rectified flow \cite{liu2022flow} further simplifies the probability path by adopting near-linear trajectories ${x}_t=(1-t){x}_0+t\epsilon$ from data to noise over continuous time $t\in[0,1]$, which induces a constant target velocity $\epsilon-x_0$.
This formulation enables efficient numerical integration and significantly reduces inference steps.
Recent models such as TangoFlux \cite{hung2024tangoflux} and MeanAudio \cite{li2025meanaudio} demonstrate the scalability of rectified flow models in the audio domain.

\begin{figure*}[t]
\centering
\includegraphics[width=1\textwidth]{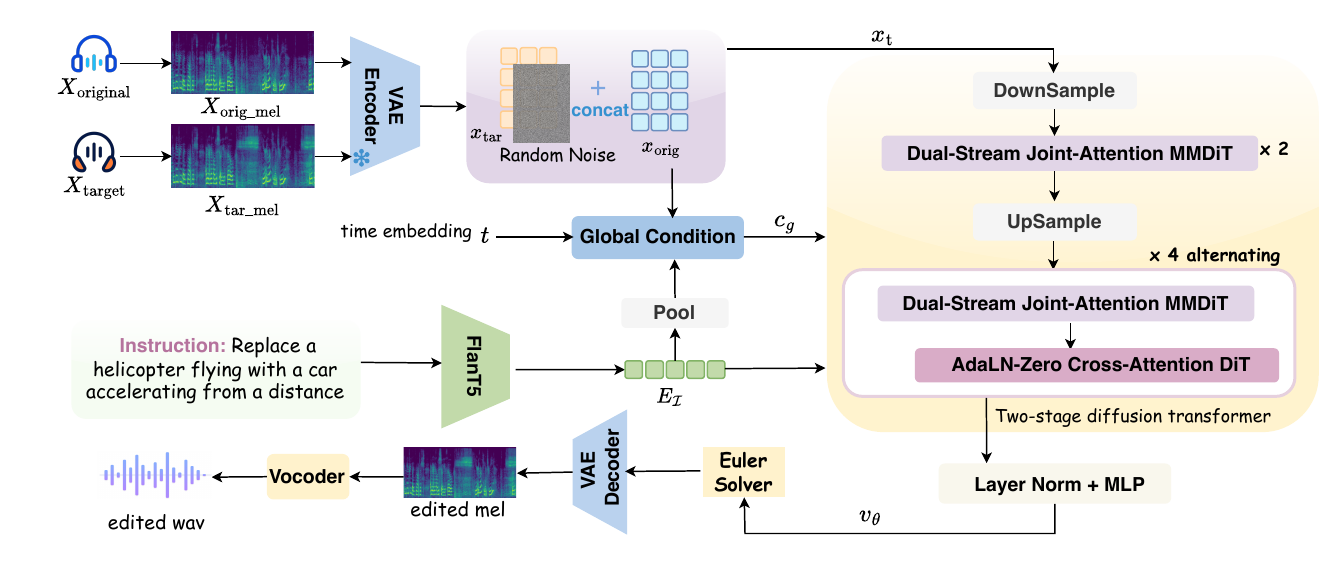}
\caption{\textbf{Overview of the RFM-Editing 2 framework.} It is a coarse-to-fine two-stage diffusion transformer framework for instruction-guided audio editing with rectified flow matching. The model first performs low-resolution audio--text semantic fusion and then refines local acoustic details at high resolution, conditioned on the editing instruction and original audio to preserve non-edited content.}
\label{fig:framework}
\end{figure*}

\subsubsection{\textbf{Training-Based Audio Editing}}
Training-based audio editing learns instruction-following capabilities from supervised data, with early methods mainly built on U-Net-based \cite{ronneberger2015u} latent diffusion models (LDMs) \cite{rombach2022high}.
AUDIT \cite{wang2023audit} trains a LDM with triplet data for instruction-guided audio editing.
Non-Rigid Prompt Editing \cite{paissan2024audio} fine-tunes a diffusion model on audio-caption pairs and performs edits via interpolation in the prompt embedding space. RFM-Editing \cite{gao2026rfm} introduces an efficient end-to-end latent diffusion text-guided audio editing framework based on RFM, learning localized velocity fields directly from instructions rather than captions or masks.

More recent methods have moved toward DiT \cite{peebles2023scalable} and MMDiT \cite{esser2024scaling} backbones.
SmartDJ \cite{lan2025guiding} adopts an audio language model initialized from Audio Flamingo 2 \cite{ghosh2025audio} that plans a sequence of atomic edits from declarative requests and executes them with a trained conditional diffusion editor for multi-step audio editing. 
MMEdit \cite{tao2025mmedit} uses a scalable synthetic pipeline and an audio language model-driven unified framework. It performs editing in the latent space by conditioning on the source audio and instruction through Qwen2-Audio \cite{chu2024qwen2}, and uses an MMDiT-based generator to achieve precise cross-modal alignment and localized multi-type editing. 
Built on the pretrained FluxAudio \cite{li2025meanaudio}, T2A-Editor \cite{zhu2026audio} performs temporally localized audio insertion and removal by injecting reference-audio and event-roll-based instructions via an external conditioning network and cross-attention.

Despite recent progress, existing training-based audio editing methods still suffer from several limitations. U-Net-based editors rely on local convolutions and often struggle with long-range audio--text semantic alignment and precise edit localization. Meanwhile, many methods rely on cross-attention for text--audio interaction to capture overall editing intent. Although DiT and MMDiT improve global modeling and multimodal fusion, current designs often rely on simple stacked blocks without exploring their multimodal interaction and conditioning mechanisms. In addition, applying joint attention over the concatenated audio and text tokens incurs quadratic complexity, making it difficult to balance editing accuracy, content preservation, and efficiency, especially for complex instructions and overlapping sound events.

\subsubsection{\textbf{Training-Free Audio Editing}}

Training-free audio editing methods typically rely on inversion inspired by \cite{mokady2023null, huberman2024edit} to recover a reproducible noise trajectory for the input audio using a pre-trained diffusion model \cite{liu2023audioldm, ghosal2023text, xue2024auffusion}. Given an input sample $x_0$, inversion progressively maps it to a sequence of noisy states $\{x_t\}_{t=1}^T$ that are consistent with the forward diffusion trajectory. At each timestep, the clean sample is estimated as:
\begin{equation}
\hat{x}_{0,\mathrm{src}}^{(t)}
=
\frac{x_t-\sqrt{1-\bar{\alpha}_t}\,\epsilon_\theta(x_t,t,c_\text{src})}
{\sqrt{\bar{\alpha}_t}},
\end{equation}
where $c_{\text{src}}$ denotes the source condition. Taking deterministic DDIM inversion as an example, the next inversion state is obtained by:
\begin{equation}
\label{equ_inversion}
x_{t+1}
=
\sqrt{\bar{\alpha}_{t+1}}\,
\hat{x}_{0,\mathrm{src}}^{(t)}
+
\sqrt{1-\bar{\alpha}_{t+1}}\,
\epsilon_\theta(x_t,t,c_{\mathrm{src}}).
\end{equation}

% After inversion, editing is performed by reverse denoising from the inverted noisy state under the target condition $c_{\text{tar}}$:
% \begin{equation}
% x_{t-1}
% =
% \sqrt{\bar{\alpha}_{t-1}}\,\hat{x}_{0,src}^{(t)}
% +
% \sqrt{1-\bar{\alpha}_{t-1}}\,\epsilon_\theta(x_t,t,c_\text{tar}).
% \end{equation}
After obtaining the inversion trajectory, reverse denoising is performed under the target condition $c_{\mathrm{tar}}$ to generate the edited audio, enabling semantic editing while preserving the non-edited source content.
Based on this paradigm, AudioEditor \cite{jia2025audioeditor} adopts DDIM inversion and null-text optimization \cite{mokady2023null} to improve reconstruction fidelity and content preservation. 
PPAE \cite{xu2024prompt} extends inversion-based editing with cross-attention map manipulation \cite{hertzprompt} during denoising to achieve more precise local editing. Zero-Shot \cite{manor2024zero} builds on DDPM inversion \cite{huberman2024edit} to extract source-corresponding noise latents, which are then reused in sampling to steer the diffusion process toward text-guided or unsupervised edits.

More recently, inversion-free editing has emerged as an alternative to inversion. AudioMorphix \cite{liang2025audiomorphix} proposes a training-free sound editor conditioned on reference audio and binary masks. SemanticAudio \cite{dai2026semanticaudio} enables attribute-level audio editing by steering the generation trajectory with the difference of velocity fields from source and target prompts, inspired by inversion-free method FlowEdit \cite{kulikov2025flowedit}. Beyond diffusion-trajectory-based methods, WavCraft \cite{liang2024wavcraft} and Audio-Agent \cite{wang2024audio} leverage large language models as planners to convert user instructions into a sequence of executable editing operations.

While training-free methods are flexible and require no labeled data, they often involve time-consuming null-text optimization in inference, which limits their practicality. Furthermore, these methods depend on full captions \cite{xu2024prompt, manor2024zero, paissan2024audio, jia2025audioeditor, dai2026semanticaudio} or modified token masks \cite{jia2025audioeditor} rather than concise editing instructions, which is time-consuming and impractical.
Due to the scarcity of high-quality audio–text pairs with detailed captions \cite{zhu2025diffusion, zhu2025zero}, we argue that an ideal audio editing system should operate from raw audio and editing instructions.

% Overall, existing audio editing methods either rely on costly inversion-based inference, require detailed captions or handcrafted masks, or adopt diffusion transformer architectures that do not fully exploit efficient conditioning mechanisms for instruction-guided editing. In contrast, our method introduces a coarse-to-fine two-stage diffusion transformer that combines efficient global semantic fusion with instruction-guided refinement. By integrating hierarchical multimodal interaction and source-aware conditioning within a rectified flow framework, the proposed approach achieves an improved trade-off between editing accuracy and computational efficiency.

In contrast, our proposed RFM-Editing 2 is a training-based text-guided audio editing framework with RFM and coarse-to-fine diffusion transformers. It combines efficient global audio--text semantic fusion at low resolution with text-guided acoustic refinement at high resolution. This design improves the balance among instruction consistency, editing precision, and computational efficiency.

\section{Proposed Method}
\label{sec:method}

\subsection{Overview}
Given an original audio clip, our goal is to edit the audio under the guidance of a textual instruction while preserving the non-edited acoustic content. As shown in Fig.~\ref{fig:framework}, we first convert the original and target waveforms into log-mel spectrograms, denoted by $X_{\text{orig\_mel}}$ and $X_{\text{tar\_mel}}$. A frozen variational autoencoder (VAE) \cite{kingma2013auto} then encodes them into latent representations $x_{\text{orig}}$ and $x_{\text{tar}}$. The editing instruction is encoded by a Flan-T5 \cite{chung2024scaling} text encoder and injected into the hybrid two-stage diffusion transformer. The transformer first performs low-resolution audio-text fusion to build coarse semantic alignment by DSJA-MMDiT blocks, and then refines local editing details with alternating DSJA-MMDiT and AZCA-DiT blocks at a high resolution. We jointly train the model on three editing tasks, including addition, removal, and replacement, in an LDM \cite{rombach2022high} based framework using the RFM objective \cite{liu2022flow} for better efficiency and stability.

\subsection{Latent Audio Editing Formulation}
During training, the source and target waveforms are resampled to 16 kHz and converted into $X_{\text{orig\_mel}}, X_{\text{tar\_mel}} \in \mathbb{R}^{T \times F}$ through STFT~\cite{allen2005unified} and mel filtering, where $T$ is the number of time frames and $F$ is the number of mel bins, typically $T=1024$ and $F=64$ used in our experiments. The VAE encoder maps each spectrogram into a latent posterior distribution $q(x \mid X) = \mathcal{E}(X)$, from which we sample $x \sim q(x \mid X)$ and obtain latent representations $x_{\text{orig}}$ and $x_{\text{tar}}$, each with a shape, typically of $8 \times 256 \times 16$. The three dimensions correspond to the latent channel dimension, the downsampled temporal frames, and the downsampled frequency bins, respectively.

The edited target latent $x_{\text{tar}}$ is used to construct the noisy latent $x_t$ and the training objective at the time step $t$ in a continuous-time RFM formulation, as in \cite{gao2026rfm}. RFM learns a continuous vector field that maps samples from a noisy distribution to the target distribution of the edited audio. Compared with standard diffusion models based on SDEs, RFM formulates a deterministic ODE process by constructing a straight interpolation path between the random Gaussian noise $\epsilon$ and the target latent $x_{\text{tar}}$ in continuous time step $t \in [0, 1]$ to obtain a perturbed latent $x_t$, eliminating the need for fine-grained time discretization:
\begin{equation}
    x_t = \left(1 - (1 - \sigma_{\min}) \cdot t\right) \cdot \epsilon + t \cdot x_{\text{tar}}, \ \epsilon \sim \mathcal{N}(0, I)
\label{addnoise}
\end{equation}
where $\sigma_{\min}$ is a small constant that keeps a minimal scale of noise near the data endpoint at $t = 1$, thereby avoiding numerical instability caused by the noise variance approaching zero. The time derivative of the interpolation path yields the ground-truth velocity field at any time step $t$:
\begin{equation}
v_{\text{target}} = \frac{d x_t}{d t} = x_{\text{tar}} - (1 - \sigma_{\min}) \cdot \epsilon. 
\label{eq:target}
\end{equation}

We provide the original latent $x_{\text{orig}}$ as an additional condition by concatenating it with the noisy latent $x_t$ along the channel dimension. This enables the model to directly access the unedited input during both training and inference, helping preserve the unchanged acoustic content while only applying edits where instructed. Meanwhile, the text instruction is encoded into the token-level embedding $E_\mathcal{I}$ by the Flan-T5.

\subsection{Coarse-to-Fine Diffusion Transformer}

We design a novel two-stage coarse-to-fine diffusion transformer architecture to balance semantic alignment, local editing precision, and efficiency. The concatenated latent $x_t \oplus x_{\mathrm{orig}}$ is patchified along the time and frequency axes and converted into a sequence of $N_a$ 512-dimensional audio tokens by a convolutional patch embedding layer. With a patch size of $2 \times 1$, the latent grid is reduced from $256 \times 16$ to $128 \times 16$, yielding $N_a=2048$ audio tokens. After adding 2D sine-cosine positional embeddings, these audio tokens are fed into the diffusion transformer together with the instruction embedding $E_\mathcal{I}$ for velocity prediction.

\textbf{Global condition modulation.}
In addition to token-level conditioning, we construct a unified global condition $c_g(t)$ for adaptive modulation following \cite{peebles2023scalable}, which combines three sources: the timestep embedding $c_t$, a pooled instruction feature $c_\mathcal{I}$, and a pooled original audio feature $c_a$, given by $c_g(t) = c_t + c_\mathcal{I} + c_a$. We inject $c_g$ into each sub-layer of all blocks through AdaLN-Zero modulation and produce a scale term $\alpha$, a shift term $\beta$, and a gate term $\gamma$, which modulate the normalized hidden states and control the residual update strength. The modulation projection is zero-initialized, so that each residual branch starts close to an identity mapping at initialization and gradually learns condition-dependent modulation. This design provides stable layer-wise control over multimodal fusion and source-aware audio refinement and improves both editing controllability and content preservation.

\begin{figure}[!t]
    \centering
    \includegraphics[width=1\columnwidth]{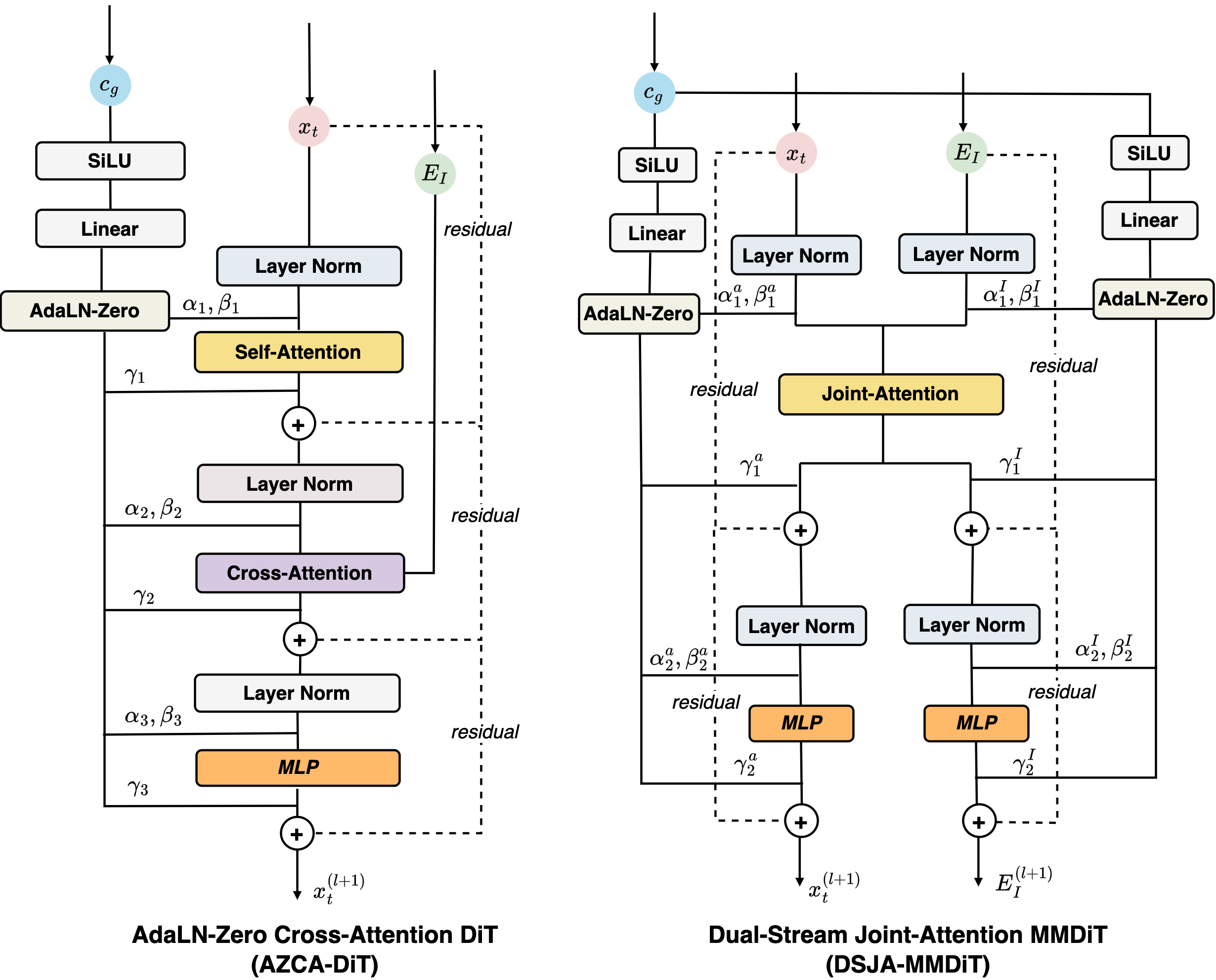}
    % \caption{Architectures of the two hybrid diffusion transformer blocks used in RFM-Editing 2: AdaLN-Zero Cross-Attention DiT (AZCA-DiT) and Dual-Stream Joint-Attention MMDiT (DSJA-MMDiT).}
    \caption{Architectures of the two hybrid diffusion transformer blocks used in RFM-Editing 2. 
\textbf{Left}: the AdaLN-Zero Cross-Attention DiT (AZCA-DiT) block for instruction-guided audio refinement. 
\textbf{Right}: the Dual-Stream Joint-Attention MMDiT (DSJA-MMDiT) block for audio--text joint fusion.}
    \label{fig:hybrid_blocks}
\end{figure}

\textbf{Low-resolution stage.}
Inspired by \cite{chen2024delta}, we perform coarse audio-text semantic alignment with two DSJA-MMDiT blocks at a lower resolution before entering fine-grained editing. The audio tokens are downsampled from $N_a$ to $\tilde{N}_a$ for low-cost global audio-text interaction. As shown in Fig. \ref{fig:hybrid_blocks}, each block follows a standard MMDiT-style design: audio and text tokens are first normalized separately, then concatenated and processed by a shared joint-attention layer, and finally split back into two streams followed by MLP updates. The joint attention complexity is reduced from $\mathcal{O}((N_a+N_\mathcal{I})^2 d)$ to $\mathcal{O}((\tilde{N}_a+N_\mathcal{I})^2 d)$ after downsampling, where $N_a$, $N_\mathcal{I}$, and $d$ denote the numbers of audio tokens, instruction tokens, and the token dimension, respectively. In our implementation, $\tilde{N}_a = N_a/4$, which lowers the quadratic attention cost while still enabling efficient global audio-instruction alignment.

\textbf{High-resolution stage.}
After the low-resolution stage, the features are upsampled back to the original resolution. The second stage then refines the editing result through four groups of alternating DSJA-MMDiT and AZCA-DiT blocks. Specifically, DSJA-MMDiT performs joint attention to further enhance audio-instruction interaction, while AZCA-DiT performs instruction-guided refinement through cross-attention, updating only the audio tokens and thus improving efficiency for fine-grained editing. This design better matches the different requirements of the refinement stage and avoids applying costly joint attention at all layers. Notably, unlike the vanilla AdaLN-Zero DiT block in \cite{peebles2023scalable} that is effective for generation, our AZCA-DiT combines token-level cross-attention with global AdaLN-Zero conditional modulation, making it better suited to controllable audio editing.

Finally, the refined audio tokens are projected, reshaped, and unpatchified back into the latent map. Conditioned on the global and token-level conditions, the model is trained to learn a continuous velocity field $v_\theta(x_t \oplus x_{\text{orig}}, c_g(t), E_{\mathcal{I}})$ that predicts the transport velocity field that drives the latent trajectory toward the target latent $x_{\text{tar}}$ by minimizing an MSE loss between the predicted $v_{\theta}$ and $v_\text{target}$:
\begin{equation}
\mathcal{L}_{\text{RFM}}
=
\mathbb{E}
\Big[
\big\|
v_{\theta}\!\left(x_t \oplus x_{\text{orig}},\, c_g(t),\, E_\mathcal{I}\right)
-
v_{\text{target}}
\big\|_2^2
\Big].
\end{equation}

\subsection{Text-Guided Audio Editing}
At inference, we leverage the trained model to perform audio editing given the input original audio and a textual instruction $\mathcal{I}$, without requiring the full target description as in \cite{jia2025audioeditor, manor2024zero}. $X_{\text{orig\_mel}}$ is first encoded by the VAE into a latent representation $x_{\text{orig}}$, while the instruction is encoded into token-level embeddings $E_{\mathcal{I}}$ using the Flan-T5 encoder.

Instead of initializing the sampling process from pure Gaussian noise when $t=0$, we adopt a flexible initialization strategy. Since audio editing aims to preserve most of the original content rather than synthesizing entirely new audio from scratch, the initial state should introduce a weak bias toward the original input to better preserve non-edited regions. Specifically, we define the starting point $x_{\text{start}}$ along the rectified interpolation path from noise $\epsilon$ to $x_{\text{orig}}$:
\begin{equation}
x_{\text{start}} = \left(1 - (1 - \sigma_{\min}) \cdot t_{\text{start}}\right) \cdot \epsilon + t_{\text{start}} \cdot x_{\text{orig}},
\label{eq:addnoise_start}
\end{equation}
where $t_{\text{start}}$ is a small adjustable parameter and we set $t_{\text{start}}=0.01$. This facilitates faithful editing by preserving non-edited regions during inference, resulting in better consistency between the edited and original audio. At each sampling step $t \in [t_{\text{start}}, 1]$, the noisy latent $x_t$ is concatenated with $x_{\text{orig}}$ along the channel dimension and passed to the trained hybrid diffusion transformer, along with $c_g$ and instruction embedding $E_{\mathcal{I}}$. The model predicts the instantaneous velocity field $v_\theta$, which is used by a continuous-time Euler solver to iteratively update the latent:
\begin{equation}
x_{t + \Delta t} = x_t + \Delta t \cdot  v_\theta(x_t \oplus x_{\text{orig}}, c_g, E_{\mathcal{I}}).
\label{eq:euler}
\end{equation}
By integrating the trajectory until $t=1$, we obtain the edited latent $\hat{x}_{\text{tar}}$, which is then decoded by the VAE decoder to reconstruct the log-mel spectrogram of the edited audio. A BigVGAN vocoder \cite{lee2022bigvgan} is used to convert the spectrogram into a waveform, producing the final edited audio output.

\section{Experiments and Results}
\subsection{Datasets and Metrics}

\begin{figure}[t]
    \centering
    \includegraphics[width=1\columnwidth]{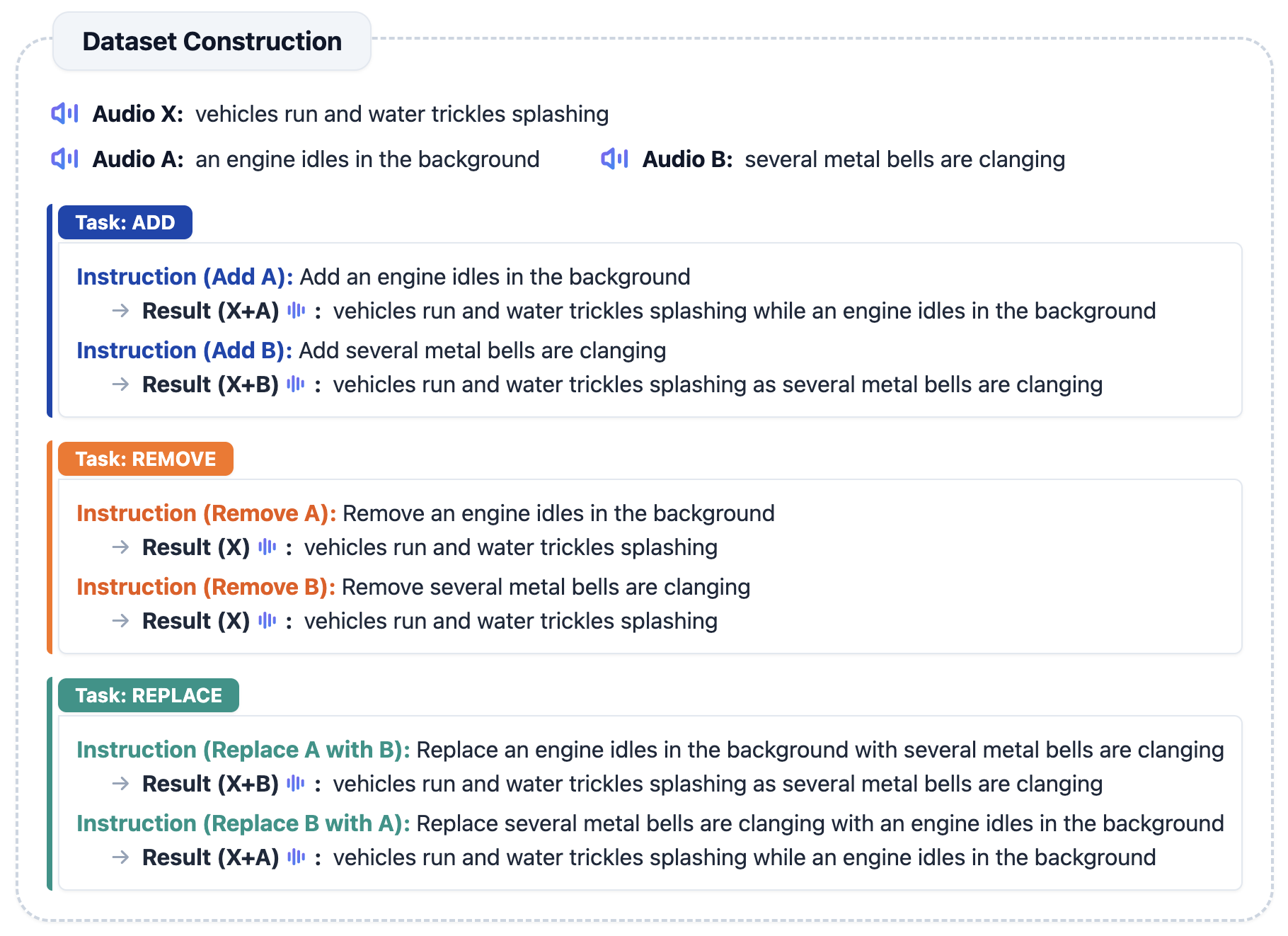}
    \caption{Dataset construction pipeline for instruction-guided audio editing. A base clip and two single-event clips are used to synthesize paired samples for the Add, Remove, and Replace tasks.}
    \label{fig:datasetconstruction}
\end{figure}

We construct instruction-based pairs using public datasets AudioCaps \cite{kim-NAACL-HLT-2019}, AudioSet \cite{gemmeke2017audio}, and AudioSetCaps \cite{bai2025audiosetcaps} for three audio editing tasks: \emph{Add}, \emph{Remove}, and \emph{Replace}. 
AudioCaps provides semantically rich captions for each clip, while AudioSet provides only class-level labels. Therefore, for the samples constructed from AudioSet, we use the captions from AudioSetCaps to enable more reliable evaluation. For AudioCaps, we retain clips containing at most three sound events as base clips, and use two randomly sampled single-event clips for mixing. In contrast, AudioSet samples use only single-category clips, thus providing a cleaner but less acoustically complex setting for comparison. All audio clips are first converted to mono, resampled to 16 kHz, and padded or trimmed to 10 seconds. As shown in Fig. \ref{fig:datasetconstruction}, given a base clip $X$ and two single-event clips $A$ and $B$, we synthesize overlapping mixtures in the waveform domain with a mixing ratio $\alpha=0.5$, optionally with additional noise $\epsilon$ weighted by $\lambda$:
\begin{equation}
\begin{aligned}
X{+}A &= \alpha X + (1-\alpha)A + \lambda \epsilon, \\
X{+}B &= \alpha X + (1-\alpha)B + \lambda \epsilon.
\end{aligned}
\end{equation}
Based on this operation, we construct three types of instruction-conditioned paired samples with the format of $\langle \text{original audio, edited audio, instruction}\rangle$: 
$\langle X, X+A, \text{Add A} \rangle$, $\langle X, X+B, \text{Add B} \rangle$, $\langle X+A, X, \text{Remove A} \rangle$, $\langle X+B, X, \text{Remove B} \rangle$, $\langle X+A, X+B, \text{Replace A with B} \rangle$, $\langle X+B, X+A, \text{Replace B with A} \rangle$.
To improve the quality of supervision, we further filter the generated samples using CLAP score, retaining only audio-text pairs with relatively consistent semantics. As shown in Table \ref{tab:dataset_summary}, we obtain two datasets with 61k and 45k paired samples, totaling 184 hours of audio, namely AudioCapsSubset and AudioSetCapsSubset. Each example has a target caption used only for evaluation.

\begin{table}[t]
\centering
\renewcommand{\arraystretch}{1.00}
\setlength{\tabcolsep}{3pt}
\caption{Summary of the datasets used in our experiments. Each audio clip is 10 s long.}
\label{tab:dataset_summary}
% \footnotesize
\renewcommand{\arraystretch}{1.08}
\setlength{\tabcolsep}{2.8pt}
\begin{tabular}{@{}llrrrr@{}}
\toprule
\textbf{Dataset} & \textbf{Task} & \textbf{Train} & \textbf{Val} & \textbf{Test} & \textbf{Duration} \\
\midrule
\multirow{3}{*}{\makecell[l]{AudioCaps\\Subset}}
& Add     & 18,041 & 2,007 & 319 & \multirow{3}{*}{111.61 h} \\
& Remove  & 18,041 & 2,007 & 360 &  \\
& Replace & 18,041 & 2,007 & 321 &  \\
\midrule
\multirow{3}{*}{\makecell[l]{AudioSetCaps\\Subset}}
& Add     & 14,039 & 1,334 & 334 & \multirow{3}{*}{72.96 h} \\
& Remove  & 14,041 & 1,333 & 333 &  \\
& Replace & 12,210 & 1,333 & 333 &  \\
\bottomrule
\end{tabular}
\end{table}

We evaluate audio editing performance from the perspectives of semantic alignment, distributional consistency, perceptual plausibility, and efficiency. CLAP Score \cite{elizalde2023clap} measures the semantic similarity of the edited audio and target caption. Fr\'echet Distance (FD) computes the distribution gap between edited and target audio using the pre-trained PANNs \cite{kong2020panns} embeddings, while Fr\'echet Audio Distance (FAD) \cite{roblek2019fr} adopts the same Fr\'echet distance formulation with VGGish \cite{hershey2017cnn} that captures low-level perceptual audio features:
% \begin{equation}
% \label{eq:unified_fd_fad}
% \mathcal{D}_{\phi}
% =
% \left\lVert \mu_g^{(\phi)} - \mu_r^{(\phi)} \right\rVert_2^2
% +
% \operatorname{Tr}\!\left(
% \Sigma_g^{(\phi)} + \Sigma_r^{(\phi)}
% - 2\left(\Sigma_g^{(\phi)} \Sigma_r^{(\phi)}\right)^{\frac{1}{2}}
% \right),
% \end{equation}
\begin{equation}
\label{eq:unified_fd_fad}
\mathcal{D}_{\phi}
=
\left\lVert \mu_g - \mu_r \right\rVert_2^2
+
\operatorname{Tr}\!\left(
\Sigma_g + \Sigma_r
- 2\left(\Sigma_g \Sigma_r\right)^{\frac{1}{2}}
\right),
\end{equation}
where $\mu_g$ and $\Sigma_g$ denote the mean and covariance of embeddings extracted from edited audio, and $\mu_r$ and $\Sigma_r$ are computed from target audio. Log-Spectral Distance (LSD) \cite{gray1976distance} is additionally used to measure the frame-wise discrepancy between edited and target audio in the log-magnitude spectrogram domain, reflecting low-level spectral fidelity.
Kullback–Leibler (KL) divergence \cite{kullback1951information} measures the divergence between the softmax-normalized class distributions of edited and target audio predicted by the PANNs classifier, reflecting the distribution consistency. Inception Score (IS) \cite{salimans2016improved} evaluates both confidence and diversity of edited samples. We also report the average editing time (AET) for each audio clip. These metrics provide a comprehensive and reproducible evaluation protocol for audio editing.

\begin{table*}[t]
\centering
\caption{Comparison testing results on the ADD, REMOVE, and REPLACE tasks of AudioCapsSubset. Values in each cell are reported in the order of \textbf{ADD / REMOVE / REPLACE}. $\uparrow$ indicates higher is better, $\downarrow$ indicates lower is better. 
The best results are shown in \textbf{bold}, and the second-best results are \underline{underlined}.}
\label{tab:audiocaps_results}
\small
\renewcommand{\arraystretch}{1.15}
\setlength{\tabcolsep}{5pt}
\resizebox{\textwidth}{!}{%
\begin{tabular}{lcccccc}
\toprule
\textbf{Method} 
& \textbf{LSD}$\downarrow$
& \textbf{CLAP}$\uparrow$ 
& \textbf{FD}$\downarrow$ 
& \textbf{FAD}$\downarrow$ 
& \textbf{KL}$\downarrow$ 
& \textbf{IS}$\uparrow$ \\
\midrule

Zero-Shot \cite{manor2024zero}
& 2.2424 / 1.9093 / 1.9611
& 0.4606 / \textbf{0.4215} / \underline{0.4209}
& 57.37 / 66.52 / 61.22
& 4.64 / 4.12 / 4.88
& 1.81 / 1.46 / 2.14
& 5.37 / 4.92 / 3.91 \\

AudioEditor \cite{jia2025audioeditor}
& 2.2238 / 2.0502 / 1.9446
& \textbf{0.4968} / 0.4079 / \textbf{0.4752}
& 31.41 / 30.46 / 31.72
& 3.61 / \textbf{2.59} / \textbf{2.73}
& 1.80 / 1.84 / 2.03
& \textbf{7.29} / \textbf{8.01} / \textbf{7.66} \\

AUDIT \cite{wang2023audit}
& 2.5776 / 3.3679 / 2.6302
& 0.1646 / 0.0004 / 0.1826
& 46.55 / 57.85 / 45.70
& 7.78 / 9.15 / 6.86
& 3.45 / 5.32 / 3.11
& \underline{5.60} / 5.97 / \underline{5.87} \\

RFM-Editing \cite{gao2026rfm}
& \underline{1.8051} / \underline{1.8567} / \underline{1.9004}
& \underline{0.4683} / 0.4086 / 0.3983
& \underline{20.58} / \underline{26.54} / \underline{29.35}
& \underline{2.94} / 2.90 / 3.43
& \underline{1.06} / \underline{1.03} / \underline{1.67}
& 5.42 / 5.89 / 4.29 \\

\midrule
\textbf{Ours}
& \textbf{1.7946} / \textbf{1.8315} / \textbf{1.8833}
& 0.4656 / \underline{0.4151} / 0.4138
& \textbf{20.54} / \textbf{25.06} / \textbf{28.77}
& \textbf{2.86} / \underline{2.83} / \underline{3.11}
& \textbf{0.97} / \textbf{0.96} / \textbf{1.52}
& 5.10 / \underline{6.02} / 4.16 \\

\bottomrule
\end{tabular}%
}
\end{table*}
\begin{table*}[t]
\centering
\caption{Comparison testing results on the ADD, REMOVE, and REPLACE tasks of AudioSetCapsSubset. Values in each cell are reported in the order of \textbf{ADD / REMOVE / REPLACE}.}
\label{tab:audioset_result}
\small
\renewcommand{\arraystretch}{1.15}
\setlength{\tabcolsep}{5pt}
\resizebox{\textwidth}{!}{%
\begin{tabular}{lcccccc}
\toprule
\textbf{Method} 
& \textbf{LSD}$\downarrow$
& \textbf{CLAP}$\uparrow$ 
& \textbf{FD}$\downarrow$ 
& \textbf{FAD}$\downarrow$ 
& \textbf{KL}$\downarrow$ 
& \textbf{IS}$\uparrow$ \\
\midrule

Zero-Shot \cite{manor2024zero}
& 2.0904 / 1.9210 / \textbf{1.8107}
& \textbf{0.5412} / \underline{0.4798} / \underline{0.5011}
& 57.49 / 62.90 / 56.87
& 1.75 / 3.90 / \textbf{1.89}
& 1.10 / \underline{1.12} / 1.29
& 3.15 / 2.79 / 2.25 \\

AudioEditor \cite{jia2025audioeditor}
& 2.5196 / 2.3252 / 1.8849
& \underline{0.5028} / \textbf{0.5019} / \textbf{0.5159}
& 34.31 / 36.69 / 31.39
& 3.57 / 4.16 / 3.19
& 1.46 / 1.35 / 1.45
& \textbf{3.55} / \textbf{3.43} / \underline{2.90} \\

AUDIT \cite{wang2023audit}
& 2.9971 / 3.6199 / 2.6371
& 0.2141 / 0.0878 / 0.2346
& 49.86 / 72.44 / 45.45
& 10.51 / 16.46 / 9.13
& 3.00 / 3.51 / 2.64
& \underline{3.46} / 2.83 / \textbf{3.18} \\

RFM-Editing \cite{gao2026rfm}
& \underline{1.8036} / \underline{1.8613} / 1.8834
& 0.4993 / 0.4257 / 0.4170
& \underline{21.51} / \underline{25.05} / \underline{25.42}
& \underline{1.65} / \underline{2.96} / 2.47
& \underline{0.83} / \textbf{0.96} / \underline{1.24}
& 2.88 / 3.16 / 2.23 \\

\midrule
\textbf{Ours}
& \textbf{1.7979} / \textbf{1.8489} / \underline{1.8724}
& 0.5027 / 0.4289 / 0.4293
& \textbf{20.65} / \textbf{24.30} / \textbf{23.98}
& \textbf{1.47} / \textbf{2.94} / \underline{2.19}
& \textbf{0.81} / \textbf{0.96} / \textbf{1.23}
& 3.02 / \underline{3.20} / 2.36 \\

\bottomrule
\end{tabular}%
}
\end{table*}

\subsection{Experimental Settings and Baselines}
 We train the model in the latent space of a pretrained VAE, where log-mel spectrograms with 1024 time frames and 64 mel-frequency bins are extracted from 10-second audio clips sampled at 16~kHz. The resulting latent representation has a spatial size of $256\times16$ and is tokenized with a patch size of $[2,1]$. Our model uses a hybrid transformer diffusion backbone with two low-resolution MMDiT blocks and four alternating groups of high-resolution MMDiT and AdaLN-Zero cross-attention DiT blocks. We adopt a velocity-based rectified flow with linear noise-to-data interpolation~\cite{liu2022flow}, where the continuous time variable $t$ is independently sampled from $\mathcal{U}(0,1)$ for each training example. Global conditioning is constructed by combining timestep, global text, and original audio features, which are injected into all transformer blocks through AdaLN-Zero modulation \cite{peebles2023scalable}. Training is conducted for 100 epochs on A100 GPUs using the AdamW \cite{loshchilov2017decoupled} optimizer with a learning rate of $5\times10^{-5}$. Classifier-free dropout is applied during training, while CFG \cite{ho2022classifier} with a guidance scale of 2.0 is used in inference. During inference, Euler integration~\cite{song2021scorebased} is used with 200 sampling steps. Validation is based on CLAP similarity of 200 samples at each epoch during training, and the checkpoint is saved according to the best CLAP score.

We compare our method against four baselines: AudioEditor \cite{jia2025audioeditor}, Zero-Shot \cite{manor2024zero}, AUDIT \cite{wang2023audit}, and RFM-Editing \cite{gao2026rfm}. AudioEditor and Zero-Shot are training-free methods based on diffusion inversion, whereas AUDIT and RFM-Editing are training-based approaches that learn instruction-guided audio editing from supervised data. We retrain the training-based editing models on our constructed dataset. These baselines represent mainstream paradigms in existing audio editing methods and provide a comprehensive basis for comparison.

\subsection{Main Results}

Tables \ref{tab:audiocaps_results} and \ref{tab:audioset_result} present the quantitative test results across three tasks (ADD, REMOVE, and REPLACE) on AudioCapsSubset and AudioSetCapsSubset, respectively. Overall, our method achieves consistently competitive performance across different tasks and datasets, indicating a balance between semantic alignment and editing fidelity. Table \ref{tab:other_metrics_main} further reports the prompt type used by each model, trainable parameter size, and average editing time on AudioCapsSubset. It shows that our method achieves competitive semantic performance with a relatively small model size and substantially higher editing efficiency than other trainable baselines, without requiring full captions.

% the the semantic alignment between edited audio and target captions measured by CLAP and the quantitative comparison of edited audio quality across three tasks (ADD, REMOVE, and REPLACE) on the test set of AudioCapsSubset and AudioSetSingleSet, evaluated using FD, FAD, KL, and IS metrics. Our method is consistently competitive and achieves lower FD/FAD/KL in most tasks, indicating improved distributional consistency of edited audio. 
% It shows that our method achieves competitive CLAP with a relatively small trainable parameter budget and roughly 2× faster editing speed than other trainable baselines under the same sampling steps without requiring the captions.

For AudioCapsSubset in Table \ref{tab:audiocaps_results}, our model achieves the best performance across the three editing tasks on most distribution-related and low-level fidelity metrics, especially LSD, FD, FAD, and KL. This indicates that the proposed hybrid diffusion transformer better preserves the global acoustic distribution and spectral fidelity while performing instruction-guided modifications. Although our method does not always obtain the highest CLAP or IS, this trade-off is reasonable for audio editing, where faithful instruction execution, acoustic consistency, and preservation of non-edited content are also critical, rather than maximizing semantic similarity or output diversity alone.

For AudioSetCapsSubset in Table \ref{tab:audioset_result}, the editing instructions are still constructed from the simple and concise class labels in AudioSet, while the captions from AudioSetCaps are used only as target captions for evaluation. This setting provides cleaner event-level supervision but still differs from AudioCaps in terms of caption richness and acoustic diversity. Even under this more challenging setting, our model remains competitive and still achieves strong distributional consistency performance in most tasks, demonstrating robustness and cross-domain generalization. In particular, our method obtains the best results on most FD, FAD, and KL metrics, and also achieves competitive LSD performance, indicating that it can preserve acoustic fidelity and distributional consistency.

The training-free method Zero-Shot \cite{manor2024zero} achieves competitive CLAP performance but lacks distributional consistency, and its results are highly sensitive to the audio sampling rate during evaluation. While AudioEditor \cite{jia2025audioeditor} attains the best CLAP and IS through local attention manipulation, its performance on global distributional consistency and fidelity, as reflected by KL and FD, remains limited. This can be attributed to the fact that training-free methods operate as an inversion-based editing framework built upon a TTA generation backbone. Since each editing operation requires triggering the full generation process, it inherently introduces higher diversity during resynthesis, leading to a higher IS. However, such aggressive resynthesis often results in unintended modifications beyond the target editing regions, which can degrade perceptual audio quality. The training-based AUDIT \cite{wang2023audit} suffers from poor performance on most metrics. RFM-Editing \cite{gao2026rfm} achieves competitive results, while our method further achieves more balanced performance across semantic alignment and fidelity.    

\begin{table}[t]
\centering
\renewcommand{\arraystretch}{1.00}
\setlength{\tabcolsep}{3pt}
\caption{Comparison of the prompt type, trainable parameter size, and editing efficiency.}
\label{tab:other_metrics_main}
\resizebox{\columnwidth}{!}{%
\begin{tabular}{lccc}
\toprule
\textbf{Method} & \textbf{Prompt} & \textbf{Trainable Params} & \textbf{AET (s) $\downarrow$} \\ 
\midrule
Zero-Shot~\cite{manor2024zero} & caption & -- & 12.11 \\
AudioEditor~\cite{jia2025audioeditor} & \begin{tabular}[c]{@{}c@{}}caption \&\\ modified tokens\end{tabular} & -- & 101.87 \\
AUDIT~\cite{wang2023audit} & instruction & 859.53M & \underline{11.00} \\
RFM-Editing~\cite{gao2026rfm} & instruction & \textbf{70.09M} & 11.23 \\
\midrule
\textbf{Ours} & instruction & \underline{78.61M} & \textbf{5.07} \\
\bottomrule
\end{tabular}%
}
\end{table}

In Table \ref{tab:other_metrics_main}, 
our method achieves the best editing efficiency, requiring only 5.07s per edit, making it substantially faster than all compared methods. Unlike training-free AudioEditor and Zero-Shot, which rely on full captions or the position of modified tokens, our model operates using concise editing instructions, similar to other training-based baselines. AudioEditor achieves a better CLAP score due to its attention replacement mechanism \cite{hertzprompt}, but its editing process is nearly an order of magnitude slower than ours due to inference-time optimization, which limits its practical usability. Our method also remains lightweight, with 78.61M trainable parameters, far fewer than AUDIT and comparable to RFM-Editing. This comparison highlights the practical advantage of our instruction-driven editing method in balancing semantic alignment, efficiency, usability, and model compactness.

\begin{figure*}[t]
\centering
\includegraphics[width=1\textwidth]{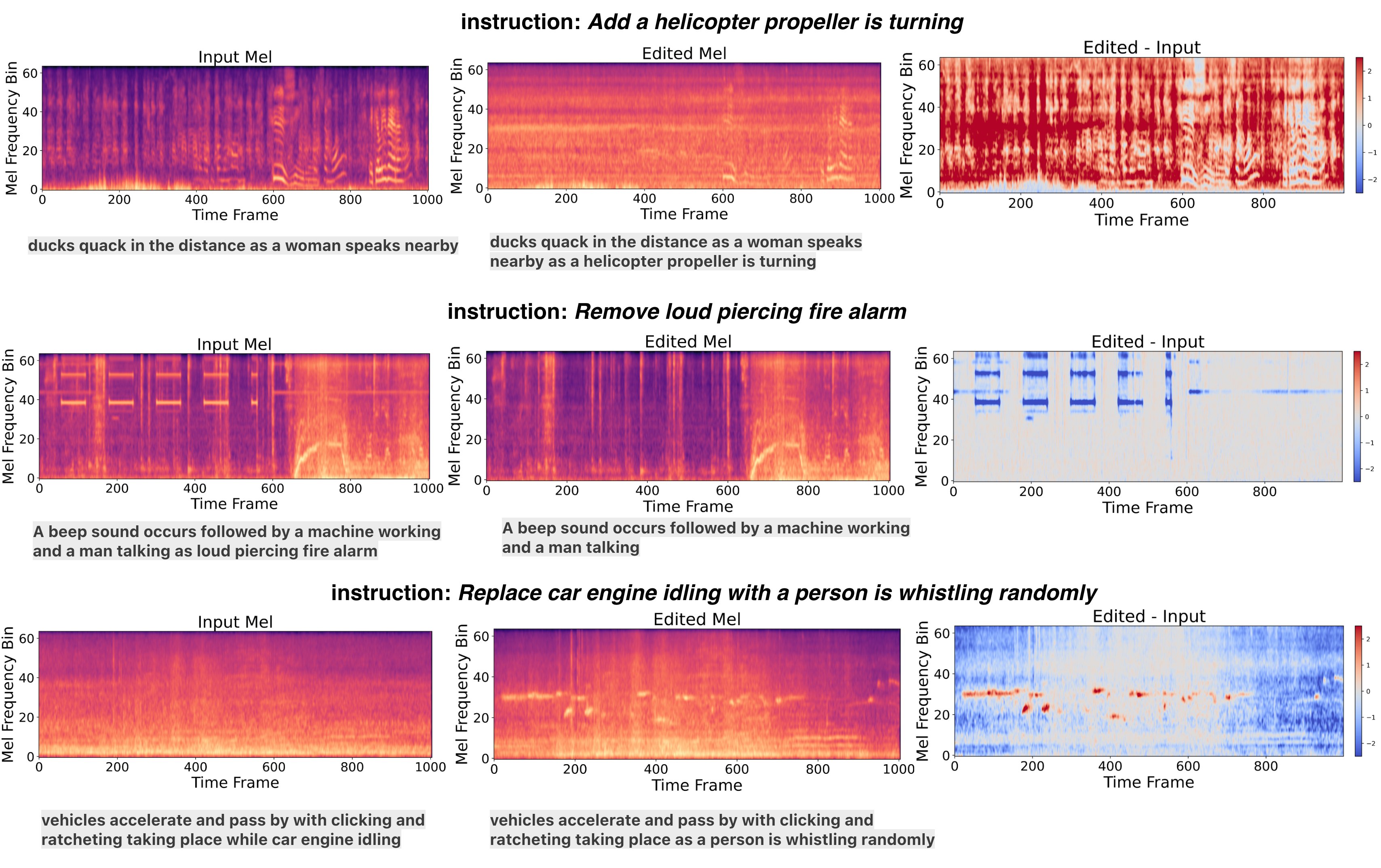}
\caption{Mel-spectrogram visualizations of Add, Remove, and Replace examples. Each row shows the input, edited output, and difference map (Edited$-$Input), where \textcolor{red}{red} regions indicate enhanced acoustic components and \textcolor{blue}{blue} regions indicate suppressed components.}
\label{fig:visual_mel}
\end{figure*}

\subsection{Ablation Study}

\begin{table}[h]
\centering
\small
\renewcommand{\arraystretch}{1.00}
\setlength{\tabcolsep}{3pt}
\caption{Ablation on the choice of diffusion transformer blocks.}
\label{tab:dit_cond_ablation}
\resizebox{\columnwidth}{!}{%
\begin{tabular}{lccccc}
\toprule
\textbf{Setting} & \textbf{CLAP$\uparrow$} & \textbf{FD$\downarrow$} & \textbf{FAD$\downarrow$} & \textbf{KL$\downarrow$} & \textbf{IS$\uparrow$} \\
\midrule
AdaLN-Zero DiTs & 0.3548 & 20.88 & 3.69 & 1.86 & \textbf{5.27$\pm$0.46} \\
Cross-attention DiTs & \textbf{0.3829} & \textbf{20.39} & \textbf{3.45} & \textbf{1.56} & 5.22$\pm$0.53 \\
In-context DiT & 0.3768 & 20.52 & 3.52 & 1.61 & 5.15$\pm$0.57 \\
\bottomrule
\end{tabular}
}
\end{table}

We conduct a systematic ablation on the module design. In Table \ref{tab:dit_cond_ablation}, we simplify the architecture to a single‑stream DiT backbone without MMDiT and compare different conditioning mechanisms. All variants are trained for 70 epochs and evaluated on the AudioCapsSubset test set. By removing the MMDiT component from the hybrid diffusion transformer, the model degenerates into a pure DiT backbone. Contrary to the common finding in image generation that AdaLN‑Zero DiT performs best, DiT with cross-attention and in-context DiT perform better in our instruction‑guided audio editing setting, with cross‑attention achieving the best overall trade‑off—highest CLAP and lowest FD/FAD/KL—indicating stronger semantic alignment and distributional consistency. The DiT with AdaLN variants show slightly higher IS but noticeably lower CLAP, suggesting that AdaLN alone is less effective at capturing fine‑grained instruction–audio interactions for editing.

\begin{table}[t]
\centering
\renewcommand{\arraystretch}{1.00}
\setlength{\tabcolsep}{3pt}
\caption{Ablation on the hybrid/alternating/multi-stage design.}
\label{tab:ablation_hybrid}
\resizebox{\columnwidth}{!}{%
\begin{tabular}{lcccccc}
\toprule
\textbf{Setting} & \textbf{CLAP$\uparrow$} & \textbf{FD$\downarrow$} & \textbf{FAD$\downarrow$} & \textbf{KL$\downarrow$} & \textbf{IS$\uparrow$} & \textbf{AET (s)$\downarrow$} \\
\midrule
w/o hybrid & 0.4023 & 17.60 & 3.11 & 1.45 & 5.76 $\pm$ 0.74 & 14.73 \\
w/o alternating & 0.4244 & 16.59 & 2.78 & 1.20 & 5.98 $\pm$ 0.62 & 5.05 \\
w/o multi-stage & 0.4255 & 16.46 & \underline{2.74} & \textbf{1.17} & \textbf{6.14 $\pm$ 0.63} & 5.57 \\
\midrule
\textbf{Ours} & \textbf{0.4257} & \textbf{16.26} & \textbf{2.62} & 1.20 & 6.04 $\pm$ 0.53 & \textbf{4.98} \\
\bottomrule
\end{tabular}%
}
\end{table}

% We ablate the hybrid architecture on AudioCapsSubset to isolate the contribution of each design component. The `w/o hybrid' setting replaces the hierarchical framework with a hybrid DiT backbone composed only of cross‑attention and in‑context DiT blocks, and yields clear drops in CLAP and distributional metrics (FD/FAD/KL) together with the slowest inference, indicating that the full hybrid design is critical for both semantic alignment and efficiency. Removing the `alternating' strategy degrades FD/FAD, suggesting that alternating joint and cross‑attention facilitates more stable distribution modeling. Dropping the `multi‑stage' low‑resolution fusion also increases FD/FAD, confirming the benefit of coarse semantic interaction before high‑resolution refinement. Our full model achieves the best or tied‑best CLAP, the lowest FD/FAD, and the fastest editing time, demonstrating that the combination of multi‑stage fusion and alternating blocks provides the most favorable trade‑off between semantic alignment, distributional consistency, and efficiency.

We ablate the hybrid architecture on AudioCapsSubset to isolate the contribution of each design component. In the `w/o hybrid' setting, the proposed dual-stream MMDiT design is simplified into a single-stream joint-attention interaction, such that the hierarchical hybrid framework is replaced by a DiT backbone composed only of cross-attention and in-context DiT blocks. This variant yields clear drops in CLAP and distributional metrics (FD/FAD/KL), indicating that the full hybrid design is important for both semantic alignment and efficiency. For the `w/o alternating' variant, the high-resolution stage uses four DSJA-MMDiT blocks followed by four AZCA-DiT blocks, instead of alternating between the two block types. Its degraded FD/FAD results suggest that alternating joint-attention fusion and cross-attention refinement helps maintain more stable distribution modeling during high-resolution generation. For the `w/o multi-stage' variant, we remove the low-resolution stage and, for a fair comparison, adopt a design with five DSJA-MMDiT blocks and five AZCA-DiT blocks operating directly in an alternating manner at high resolution. The resulting increase in FD/FAD confirms the benefit of introducing coarse semantic interaction before high-resolution refinement. Overall, our full model achieves the best CLAP, the lowest FD/FAD, and the fastest editing time, demonstrating that the combination of multi-stage fusion and alternating block design provides the most favorable trade-off between semantic alignment, distributional consistency, and efficiency.

\subsection{Visualization}
The visualizations in Fig. \ref{fig:visual_mel} provide an intuitive view of how our model performs Add, Remove, and Replace operations in the Mel-spectrogram domain. In the Add example, the edited spectrogram shows newly introduced, relatively continuous energy patterns, especially in the low-to-mid frequency bands, which are consistent with the rotating helicopter propeller sound. The corresponding difference spectrogram contains large red regions, indicating that the model primarily injects new acoustic components rather than removing existing ones. In the Remove example, the input spectrogram contains clear periodic high-energy structures corresponding to the loud fire alarm. After editing, these repetitive patterns are largely suppressed, while the remaining structures are relatively preserved. This is also reflected in the difference spectrogram, where blue regions appear at the alarm locations, showing targeted energy reduction. In the Replace example, the relatively steady low-frequency energy of the idling car engine is reduced, while new sparse and irregular patterns appear in the edited spectrogram, consistent with randomly occurring whistling. The red blocks in the difference spectrogram highlight where these new whistling-related components are introduced, while the blue regions indicate the removal of the original engine sound. Overall, these visualizations demonstrate that the model performs structured and semantically guided modifications in accordance with the textual instruction while preserving unrelated content, including addition, suppression, and replacement of acoustic events, as reflected by the localized changes in the difference spectrogram.

\subsection{Limitation}

Despite achieving satisfactory performance, our method still has several limitations.

\textit{1) Fixed Audio Length:} The current architecture is primarily designed for 10-second audio clips. Although longer audio could be handled by segmenting the input or adapting the temporal resolution of latent representations, these extensions may introduce boundary artifacts and degrade editing performance. Improving the model's generalization to variable-length audio remains an important direction for future work.

\textit{2) Synthetic Training Data:} Our training data is constructed from synthetic audio mixtures. However, such synthetic construction may not fully reflect the complexity of real-world audio editing, where sound events can be highly dynamic and affected by background noise, reverberation, recording conditions, and complex source interactions. More diverse and realistic editing datasets are needed to improve robustness in practical scenarios.

\textit{3) Lack of Explicit Temporal Control:} The model performs instruction-level editing without explicit temporal annotations. This limits its applicability to scenarios that require fine-grained temporal editing, such as modifying only a specific occurrence of a repeated sound event.

\textit{4) CLAP-based Model Selection:} During validation, the best model is selected mainly according to CLAP similarity. Although CLAP is effective for measuring global text--audio semantic alignment, it may not fully capture localized editing accuracy, background preservation, or perceptual audio quality. Future work could adopt multi-objective validation criteria that jointly consider semantic alignment, edit correctness, perceptual quality, and preservation of non-target audio content.

\section{Conclusion}
This paper presented RFM-Editing 2, a novel framework for text-guided audio editing based on rectified flow matching and coarse-to-fine diffusion transformer. The proposed framework adopts a two-stage hybrid architecture, where low-resolution dual-stream joint-attention MMDiT blocks establish efficient audio--text semantic fusion and high-resolution alternating DSJA-MMDiT and AdaLN-Zero cross-attention DiT blocks refine local editing details. By combining global AdaLN-Zero modulation with token-level instruction and source-audio conditioning, the model achieves effective control over editing behavior while preserving non-edited content. Experimental results show that the proposed method achieves a favorable trade-off among semantic alignment, distributional consistency, and inference efficiency, outperforming or remaining competitive with representative training-free and training-based baselines. The ablation results further verify the importance of the hybrid architecture, the alternating refinement strategy, and the multi-stage design. Future work will explore more flexible real-world audio editing settings, including more complex acoustic scenarios and broader open-domain instruction following.

% or


% \appendices

% \section{}

\section*{Acknowledgment}
This work was supported by a research scholarship from the China Scholarship Council (CSC) and a scholarship from the Centre for Vision, Speech and Signal Processing, University of Surrey.


\bibliographystyle{IEEEtran}
\bibliography{reference}


% \end{IEEEbiography}


%\vfill

\end{document}